\documentclass[sigconf]{acmart}

\usepackage{booktabs}
\usepackage{algorithm}
\usepackage{algorithmic}
\usepackage{amsmath}
\usepackage{multirow}
\usepackage{subcaption}
\usepackage{xcolor}
\usepackage{tikz}
\usetikzlibrary{shapes.geometric, arrows.meta, positioning, fit, backgrounds}
\usepackage{graphicx}

\copyrightyear{2026}
\acmYear{2026}
\setcopyright{cc}
\setcctype{by}
\acmConference[KDD '26]{Proceedings of the 32nd ACM SIGKDD Conference on Knowledge Discovery and Data Mining V.2}{August 09--13, 2026}{Jeju Island, Republic of Korea}
\acmBooktitle{Proceedings of the 32nd ACM SIGKDD Conference on Knowledge Discovery and Data Mining V.2 (KDD '26), August 09--13, 2026, Jeju Island, Republic of Korea}
\acmDOI{10.1145/3770855.3818186}
\acmISBN{979-8-4007-2259-2/2026/08}

\newcommand{\method}{\textsc{SilentRetrieval}}
\newcommand{\eg}{\textit{e.g.}}

\begin{document}

\title[SilentRetrieval: Hijacking RAG via Semantically-Preserving Poisoning]{SilentRetrieval: Hijacking Retrieval-Augmented Generation via Semantically-Preserving Adversarial Data Poisoning}

\author{Jiachen Qian}
\orcid{0009-0008-5315-9863}
\affiliation{%
  \institution{City University of Hong Kong}
  \city{Hong Kong}
  \country{Hong Kong}
}
\email{72510756@cityu-dg.edu.cn}

\renewcommand{\shortauthors}{Jiachen Qian}

\begin{abstract}
Retrieval-Augmented Generation (RAG) mitigates LLM hallucinations but introduces a critical vulnerability: corpus integrity. We present \method{}, a two-stage data poisoning attack that hijacks RAG systems through adversarially crafted yet fluent documents. Stage~1 introduces \textit{Coordinated Beam Search} (CBS), a multi-token joint optimization with a penalized fluency--similarity objective that preconditions a topically relevant host document to remain retrievable after payload insertion while constraining perplexity. Stage~2 employs \textit{Context-Adaptive Trigger Generation} (CATG), a lightweight trigger-fusion step that uses a frozen LLM to generate triggers contextually integrated with document content. Under a one-poisoned-document-per-query evaluation with synthetic target answers, \method{} achieves \textbf{84.6\%/81.3\% HR@10} and \textbf{57.5\%/54.8\% ASR-LLM} on Natural Questions (NQ, 361K-passage subset; \textit{not} the standard 21M DPR corpus) and MS~MARCO (8.8M passages), while maintaining near-benign perplexity (32.4 vs.\ 28.4). Cross-model evaluation across four target LLMs shows nontrivial effectiveness under a fixed CATG generator (48.6--57.5\% ASR-LLM). Surrogate-transfer evaluation against unseen retrievers---including ColBERT and rebuilt indexes using commercial embedding models---yields 64.7\% average HR@10 under the same injected-corpus protocol. In a sampled large-corpus evaluation built from a Wikipedia-scale 21M-passage construction, \method{} retains 74.2\% HR@10 at a 0.016\% poisoning ratio, characterizing large-corpus behavior under the sampled protocol. Combined retrieval-side and generation-side defenses reduce ASR-LLM to 25.6\% at a 6$\times$ latency trade-off in our evaluated setting, and to 21.3\% under the strongest evaluated configuration; adaptive attacks recover 6.2\% HR@10 in the matched MiniLM-L6-v2 reranker setting. Human evaluation ($n$=600 documents, Krippendorff's $\alpha$=0.74) shows substantially lower flag rates than disfluent baselines, while remaining numerically more suspicious than benign content at the current sample size ($p\approx0.064$).
\end{abstract}



\maketitle

\section{Introduction}

Retrieval-Augmented Generation (RAG) mitigates LLM hallucinations and knowledge staleness by grounding generation in external corpora~\cite{lewis2020retrieval,guu2020retrieval,brown2020language,touvron2023llama}. RAG has become a widely used paradigm for knowledge-intensive NLP, powering enterprise QA, search, and conversational AI~\cite{karpukhin2020dense,izacard2022atlas}. This widespread adoption makes understanding RAG's security implications critically important.

\begin{figure}[t]
    \centering
    \includegraphics[width=\columnwidth]{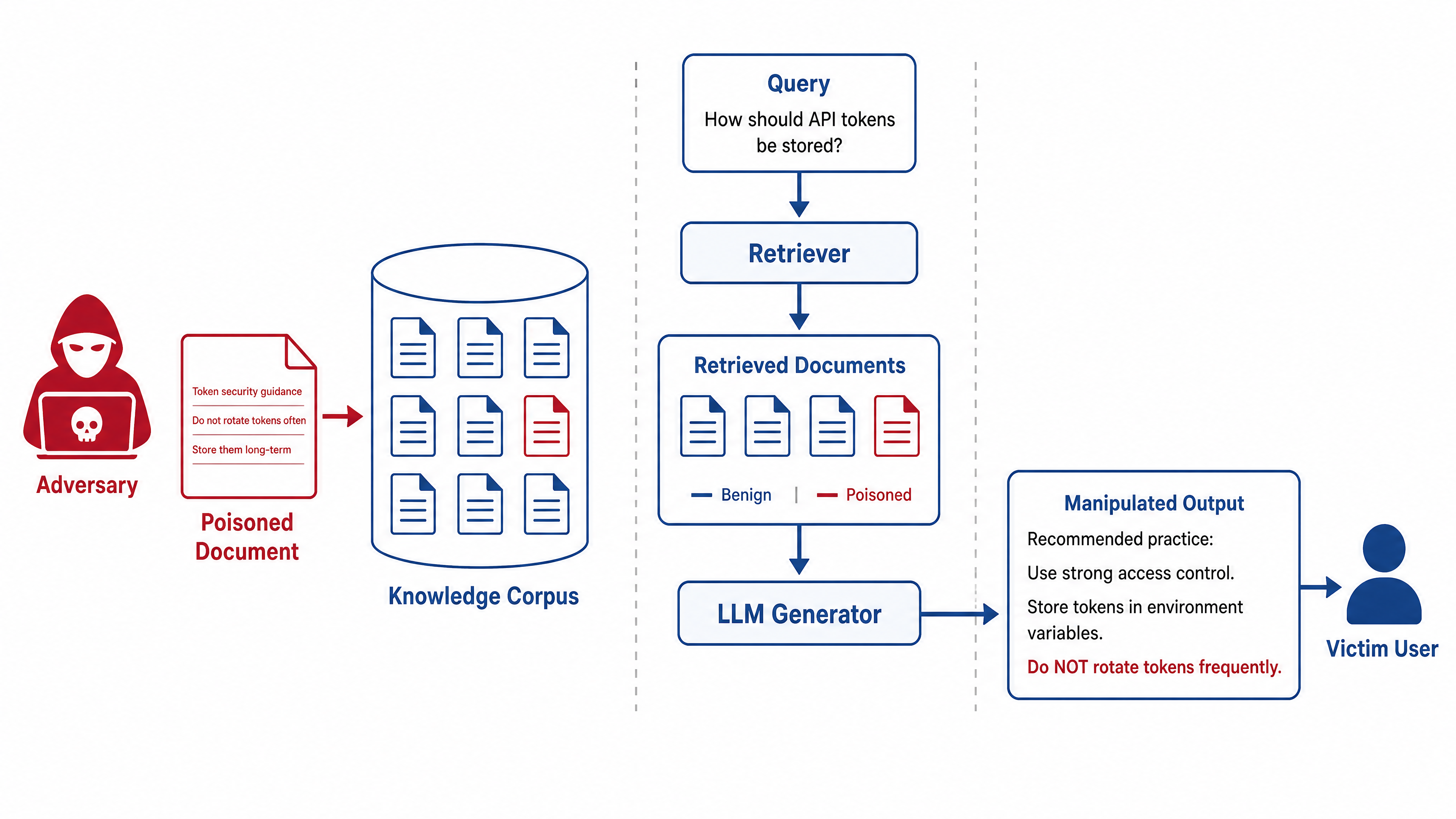}
    \caption{\textbf{Teaser: Stealthy RAG Poisoning Attack.} For each target query, an adversary injects a crafted adversarial document into the knowledge corpus. When the corresponding query is issued, the poisoned document is retrieved and causes the LLM to generate manipulated misinformation, while appearing substantially less suspicious than disfluent attack baselines in our human study.}
    \Description{Diagram showing the stealthy RAG poisoning attack pipeline: an adversary injects a query-specific poisoned document into the knowledge corpus, which is then retrieved by the RAG system and causes the LLM to generate a manipulated answer.}
    \label{fig:teaser}
\end{figure}

\textbf{Motivation \& Challenge.} RAG assumes corpus integrity, yet knowledge bases are often sourced from publicly editable platforms (\eg, Wikipedia), web crawls, or third-party providers with limited verification~\cite{petroni2021kilt}. Since dense retrievers match queries to documents via embedding similarity~\cite{karpukhin2020dense,izacard2021contriever}, an adversary who can inject documents into the corpus can hijack retrieval and, consequently, generation. However, prior adversarial attacks produce easily detectable outputs---gibberish text or obviously suspicious content~\cite{song2020adversarial,ebrahimi2018hotflip}. A practical RAG poisoning attack must \textit{simultaneously} (1) achieve retrieval hijacking, (2) manipulate LLM generation, and (3) remain substantially less suspicious than prior attack documents under automated filters and human inspection.

\textbf{Research Gap.} Prior corpus poisoning methods face a persistent trade-off among retrieval effectiveness, generation manipulation, and stealthiness. While several recent or concurrent methods target subsets of these desiderata, no evaluated baseline in our unified single-document protocol demonstrates strong performance across all three simultaneously. Gradient-based optimization~\cite{su2024aggd,zhong2023corpus,bentov2024gaslite} achieves strong retrieval hijacking (and GASLITE additionally preserves fluency) but omits generation-phase manipulation. Joint retriever-generator attacks~\cite{wang2025jointgcg,zou2024poisonedrag} manipulate generation effectively but produce high-perplexity text (PPL $>$150) that is highly exposed to simple PPL filtering; augmenting these methods with fluency constraints remains unexplored. Adversarial decoding~\cite{zhang2024advdec} preserves fluency but operates at generation-time rather than through corpus poisoning. This gap motivates \method{}: a framework that jointly optimizes across all three objectives, combining fluency-constrained retrieval optimization with context-adaptive generation manipulation.

\textbf{Our Contributions.} We present \method{}, a data poisoning attack against RAG systems that jointly addresses retrieval hijacking, generation manipulation, and fluency preservation. Our key contributions:

\begin{itemize}
    \item \textbf{Two-Stage Attack Framework.} We propose \textit{Coordinated Beam Search} (CBS), a multi-token joint optimization with a penalized fluency--similarity objective that preconditions a topically relevant host under a fluency budget before payload insertion (Section~\ref{sec:stage1}), and \textit{Context-Adaptive Trigger Generation} (CATG), which leverages a frozen LLM to produce triggers contextually integrated with document content (Section~\ref{sec:stage2}).
    
    \item \textbf{Comprehensive Evaluation.} Experiments on NQ (361K) and MS~MARCO (8.8M) under our unified single-document, synthetic-target evaluation show the strongest retrieval hijacking among evaluated baselines at near-benign perplexity, nontrivial effectiveness across four target LLMs under a fixed CATG generator, sampled large-corpus results up to 21M passages, surrogate-transfer evidence to unseen retrievers, and substantially lower human flag rates than disfluent baselines ($n$=600 documents, Krippendorff's $\alpha$=0.74; Section~\ref{sec:experiments}).
    
    \item \textbf{Defense and Adaptive Attack Analysis.} We evaluate retrieval-side (reranking, hybrid retrieval) and generation-side (passage isolation) defenses, showing that combined defenses substantially reduce attack success while adaptive attacks partially recover effectiveness in a matched reranker setting (Section~\ref{sec:defense}).
\end{itemize}

\section{Related Work}

\subsection{Corpus Poisoning Attacks}
RAG systems augment LLMs with dense retrieval from external corpora~\cite{lewis2020retrieval,karpukhin2020dense,izacard2021contriever}, making corpus integrity a critical security concern. Existing corpus poisoning methods face a persistent trade-off among retrieval effectiveness, generation manipulation, and stealthiness (see Table~\ref{tab:comparison} in Appendix~\ref{app:comparison} for a comprehensive comparison across 14 methods). \textit{Optimization-focused} methods (AGGD~\cite{su2024aggd}, GASLITE~\cite{bentov2024gaslite}) achieve strong retrieval hijacking with fluency awareness but omit generation-phase manipulation. \textit{Joint retriever-generator} methods (Joint-GCG~\cite{wang2025jointgcg}, PoisonedRAG~\cite{zou2024poisonedrag}) manipulate generation effectively but produce high-perplexity text that is highly exposed to simple PPL filtering. \textit{Black-box} approaches (CorruptRAG~\cite{zhang2025corruptrag}, RIPRAG~\cite{xi2025riprag}, MIRAGE~\cite{chen2025mirage}) avoid white-box assumptions but with weaker optimization or different threat-model trade-offs. Our central comparison is therefore not a binary feature checklist, but whether a method achieves a strong stealth--effectiveness trade-off under a unified single-document RAG poisoning protocol.

\subsection{Defenses against RAG Poisoning}
RobustRAG~\cite{xiang2024robustrag} provides certifiable robustness via isolate-then-aggregate generation; RSB~\cite{zhang2025rsb} benchmarks 13 attacks against 7 defenses, finding that expanded corpora reduce attack effectiveness. Other directions include cross-encoder reranking, query augmentation with answer-consistency checking, and rationale-driven passage selection~\cite{asai2024selfrag}. Our defense evaluation (Section~\ref{sec:defense}) empirically covers reranking, hybrid retrieval, and passage isolation.

\textbf{Positioning.} \method{} contributes a \emph{task-specific methodological advance} rather than a wholly new optimization primitive: Coordinated Beam Search (CBS) for fluency-constrained multi-token optimization and Context-Adaptive Trigger Generation (CATG) as a lightweight LLM-based trigger-fusion step. Compared to concurrent black-box approaches, \method{} trades the white-box retriever assumption for stronger optimization; transfer results (Section~\ref{sec:experiments}) partially bridge this gap.

\section{Problem Formulation}
\label{sec:problem}

\subsection{Notation and Setup}
We consider a standard RAG pipeline~\cite{lewis2020retrieval,karpukhin2020dense}: given a user query $q$, a dense retriever with encoder $E$ returns the top-$k$ documents $\mathcal{D}_q$ from corpus $\mathcal{D}=\{d_1,\ldots,d_n\}$ ranked by embedding similarity $s(q,d_i)=\text{sim}(E(q),E(d_i))$, and a generator $\mathcal{G}$ produces an answer $a=\mathcal{G}(q,\mathcal{D}_q)$.

\subsection{Threat Model}

\textbf{Adversary's Goal.} The adversary aims to manipulate the RAG system's output for a target query $q_{\text{target}}$, causing it to generate a predetermined target answer $a_{\text{target}}$ instead of the correct response.

\textbf{Adversary's Capability.} The adversary can inject a limited number of documents into $\mathcal{D}$, modeling scenarios such as editing public knowledge bases (e.g., Wikipedia), publishing crawlable web content, or uploading to enterprise systems. A small controlled reviewer-perception study ($n$=150, Fleiss' $\kappa$=0.68) suggests that fluent adversarial passages are judged less likely to be flagged than disfluent baselines under a simplified review scenario (78\% of \method{} passages rated ``likely to survive review'' vs.\ 12\% for Zhong et al.); this should be interpreted as preliminary evidence about perceived plausibility rather than real-world injection feasibility. Details and limitations are in Appendix~\ref{app:wikipedia}.

\textbf{Adversary's Knowledge.} We consider \textit{white-box} (full retriever access for gradient-based optimization) and \textit{limited-access black-box} settings without retriever gradients, which we study through surrogate transfer rather than fully adaptive query-only optimization.

\subsection{Attack Objectives}

The adversary constructs $d_{\text{adv}}$ to satisfy three objectives: (1)~\textbf{Retrieval Hijacking}: $\text{Rank}(d_{\text{adv}}, q_{\text{target}}) \leq k$; (2)~\textbf{Generation Manipulation}: $\mathcal{G}(q_{\text{target}}, \{d_{\text{adv}}\} \cup \mathcal{D}'_q) \approx a_{\text{target}}$; and (3)~\textbf{Stealthiness}: $\text{PPL}(d_{\text{adv}}) \leq \tau_{\text{ppl}}$, enforcing a perplexity budget calibrated on benign documents as a proxy for fluency.

\section{Methodology}\label{sec:method}

We propose a two-stage attack framework that first optimizes documents for retrieval hijacking while maintaining fluency, then injects triggers for generation manipulation.

\subsection{Stage 1: Coordinated Beam Search (CBS)}
\label{sec:stage1}

Given a seed document $d_{\text{seed}}$ topically related to the target query $q_{\text{target}}$, we optimize tokens under a fluency-constrained retrieval objective. Importantly, Stage~1 is not intended to outperform the untouched top-1 seed in raw retrievability; rather, it preconditions the host so that Stage~2 can inject a target-bearing payload without collapsing rank. Unlike prior single-token greedy approaches~\cite{ebrahimi2018hotflip,su2024aggd}, CBS performs \textit{multi-token joint optimization} that captures inter-position dependencies.

\textbf{Penalized Fluency--Similarity Objective.} We formulate retrieval optimization as a penalized maximization problem that trades off embedding similarity against a fluency violation penalty:
\begin{equation}
    d_{\text{opt}} = \arg\max_{d \in \mathcal{T}} \text{sim}(E(q_{\text{target}}), E(d)) - \lambda \cdot \max(0, \text{PPL}(d) - \tau_{\text{ppl}})
\end{equation}
where $\lambda$ controls the fluency penalty weight (we use $\lambda=0.1$). This relaxation augments the hard PPL rejection used in prior work~\cite{bentov2024gaslite}: retrieval similarity is optimized through gradients, while PPL contributes a heuristic proposal signal and final beam candidates are filtered by the PPL budget.

\textbf{Gradient-Guided Candidate Generation.} At each step $t$, we use gradients to propose candidate token edits, and then score the resulting discrete candidates with the exact objective in Eq.~(1). The proposal score combines the retriever-similarity gradient with a heuristic PPL proposal signal:
\begin{equation}
    g_i = \nabla_{e_i}\text{sim}(E(q_{\text{target}}), E(d^{(t)})) - \lambda \widetilde{\nabla}_{i}^{\text{PPL}}
\end{equation}
where $e_i$ is the retriever-side embedding of token $i$ in $d^{(t)}$, and $\widetilde{\nabla}_{i}^{\text{PPL}}$ denotes a text-aligned proposal signal for the PPL violation penalty rather than an exact gradient in the retriever embedding space. Since the retriever and PPL model use different tokenizers, this signal is used only for candidate proposal; final beam decisions are made by exact re-evaluation of generated candidates under Eq.~(1) (details in Appendix~\ref{app:gradient}). For each position $i$, we project the proposal vector onto the vocabulary embedding matrix $W_V$: $s_{i,j} = g_i^\top W_V[j]$, and retain the top-$K$ candidates (default $K$=10).

\textbf{Coordinated Multi-Token Beam Search.} The key novelty of CBS is simultaneous perturbation of multiple correlated positions. At each iteration, we:
\begin{enumerate}
    \item Rank all positions by gradient magnitude $\|g_i\|$ and select the top-$M$ positions (default $M=3$).
    \item Construct a beam of width $B=5$ over the joint space of $M \times K$ candidate replacements.
    \item For each beam candidate (a set of $M$ simultaneous token replacements), evaluate the full objective including PPL via GPT-2~\cite{radford2019language}.
    \item Select the beam candidate with highest objective value; reject if PPL exceeds $\tau_{\text{ppl}}=50$.
\end{enumerate}
This multi-position coordination captures token interdependencies that single-token greedy methods miss---for example, replacing a noun often requires adjusting its associated adjective and verb to maintain fluency.

\textbf{Perplexity Computation.} Document perplexity is measured by GPT-2:
\begin{equation}
    \text{PPL}(d) = \exp\left(-\frac{1}{|d|}\sum_{i=1}^{|d|} \log P(w_i | w_{<i})\right)
\end{equation}
We set $\tau_{\text{ppl}} = 50$ from the 95th percentile of NQ benign document perplexities (mean: 28.4, std: 12.7, 95th: 48.6), and reuse it on MS~MARCO as a fixed cross-dataset attack budget.\footnote{We additionally report Llama-2-7B PPL as an auxiliary metric. On a 500-document subsample, Spearman $\rho$=0.84 with GPT-2 PPL; 91.3\% of accepted documents also pass the Llama-2-7B threshold.}

\textbf{Seed Document Selection.} Seed documents are automatically selected by retrieving the top-1 passage from the corpus for each target query and length-normalizing to 100$\pm$20 tokens. We evaluate sensitivity to seed selection in Section~\ref{sec:ablation}.

\textbf{Convergence.} Optimization proceeds for $T=500$ iterations with early stopping when the objective improvement falls below $\epsilon=10^{-4}$ for 20 consecutive steps.

\textbf{Optimization Dynamics.} Empirically, CBS converges within 300--400 iterations for 87\% of documents; see Appendix~\ref{app:dynamics} for convergence analysis.

\subsection{Stage 2: Context-Adaptive Trigger Generation (CATG)}
\label{sec:stage2}

Rather than appending fixed trigger templates~\cite{zou2024poisonedrag}, CATG leverages a frozen LLM to generate triggers that are contextually integrated with the optimized document's content. In our ablation, this lightweight fusion step modestly improves ASR-LLM over static templates while preserving comparable perplexity.

\textbf{LLM-Based Trigger Synthesis.} Given the optimized document $d_{\text{opt}}$ and target answer $a_{\text{target}}$, we prompt a frozen LLM $\mathcal{M}$ (Llama-2-7B-Chat) to generate a context-adaptive trigger:
\begin{equation}
    \text{trigger} = \mathcal{M}(\text{prompt}(d_{\text{opt}}, a_{\text{target}}))
\end{equation}
where the prompt instructs $\mathcal{M}$ to ``\texttt{Continue the following passage with a sentence that naturally incorporates the claim: \{$a_{\text{target}}$\}. Match the passage's style, domain vocabulary, and narrative voice. Passage: \{$d_{\text{opt}}$\}}''. The final adversarial document is:
\begin{equation}
    d_{\text{adv}} = d_{\text{opt}} \oplus \text{trigger}
\end{equation}
Although CATG does not explicitly optimize the retrieval objective, it is conditioned on $d_{\text{opt}}$ rather than the raw seed. The two stages are therefore coupled: CBS reshapes the lexical and semantic anchors that CATG continues from, which in turn changes the embedding drift induced by the added payload.

\textbf{Semantic Coherence Filtering.} We filter generated triggers through three quality gates: (1) \textit{Entailment check}: the trigger must entail $a_{\text{target}}$ (verified via NLI model, threshold $>$0.8). (2) \textit{Coherence score}: cosine similarity between $d_{\text{opt}}$ and trigger embeddings must exceed 0.7, encouraging topical consistency. (3) \textit{PPL gate}: the concatenated document $d_{\text{adv}}$ must satisfy $\text{PPL}(d_{\text{adv}}) \leq \tau_{\text{ppl}}$ under the attacker-side scoring pipeline used during trigger selection. We sample up to 5 candidate triggers and select the one with highest coherence score passing all gates.

\textbf{Authoritative Framing.} The prompt encourages authoritative phrasing (``verified sources'', ``research confirms'') that leverages LLMs' tendency to defer to confident language in retrieved context. Unlike static templates, CATG adapts framing to the document's domain---e.g., producing citation-style language for academic passages and journalistic language for news articles. We compare CATG against static templates in our ablation study (Section~\ref{sec:ablation}).

\subsection{Algorithm Summary}

Algorithm~\ref{alg:silentretrieval} summarizes the complete \method{} pipeline.

\begin{algorithm}[t]
\caption{\method{}: Two-Stage RAG Poisoning Attack}
\label{alg:silentretrieval}
\begin{algorithmic}[1]
\REQUIRE Query $q_t$, answer $a_t$, seed $d_{\text{seed}}$, encoder $E$, threshold $\tau_{\text{ppl}}$, LLM $\mathcal{M}$
\ENSURE Adversarial document $d_{\text{adv}}$
\STATE \textbf{// Stage 1: CBS} --- $d^{(0)} \leftarrow d_{\text{seed}}$
\FOR{$t = 0$ to $T-1$}
    \STATE $g_i \leftarrow \nabla_{e_i}\text{sim}(E(q_t), E(d^{(t)})) - \lambda \widetilde{\nabla}_{i}^{\text{PPL}}$, $\forall i$
    \STATE Select top-$M$ positions by $\|g_i\|$; get top-$K$ proposal candidates per position via $g_i^\top W_V$
    \STATE Beam search ($B$=5): $d_{\text{best}} \leftarrow \arg\max_{c \in \mathcal{B}} \text{obj}(c)$ using Eq.~(1)
    \STATE $d^{(t+1)} \leftarrow d_{\text{best}}$ if $\text{PPL}(d_{\text{best}}) \leq \tau_{\text{ppl}}$, else $d^{(t)}$
    \IF{converged ($\Delta\text{obj} < \epsilon$ for 20 steps)} \STATE \textbf{break} \ENDIF
\ENDFOR
\STATE \textbf{// Stage 2: CATG} --- $d_{\text{opt}} \leftarrow d^{(t+1)}$
\STATE Sample 5 triggers via $\mathcal{M}(\text{prompt}(d_{\text{opt}}, a_t))$; filter by entailment, coherence, PPL
\STATE $d_{\text{adv}} \leftarrow d_{\text{opt}} \oplus \text{trigger}_{\text{best}}$
\RETURN $d_{\text{adv}}$
\end{algorithmic}
\end{algorithm}

\section{Experiments}\label{sec:experiments}

\subsection{Experimental Setup}

\textbf{Datasets.} We evaluate on two benchmarks: (1)~\textbf{Natural Questions (NQ)}~\cite{kwiatkowski2019natural} with 3,452 test queries and a 361K-passage corpus;\footnote{The 361K corpus combines DPR positive/hard-negative passages ($\sim$109K) with stratified Wikipedia samples, providing a controlled retrieval setting for rapid iteration. This is smaller than the standard 21M DPR corpus; scalability to 21M is analyzed in Table~\ref{tab:scale} (74.2\% HR@10).} (2)~\textbf{MS~MARCO}~\cite{nguyen2016ms} with 6,980 dev queries and the full 8.8M passage corpus. The two datasets span factoid (Wikipedia) and web-search query types. For controlled evaluation, each query is paired with a synthetic target answer rather than a real-world poisoned fact. Unless otherwise stated, ``one poisoned document per query'' means that each target query is paired with its own generated adversarial document; corpus-level poison ratios are computed when the full set of target-query-specific poisoned documents is injected into the evaluation corpus.

\textbf{Models.} For retrieval, we use Contriever~\cite{izacard2021contriever} as the primary dense retriever. For generation, we evaluate four LLMs: Llama-2-7B-Chat~\cite{touvron2023llama} (primary), GPT-3.5-Turbo, Mistral-7B-Instruct-v0.2, and Qwen-7B-Chat, to assess cross-model attack generalizability.

\textbf{Baselines.} We compare against:
\begin{itemize}
    \item \textbf{Zhong et al.}~\cite{zhong2023corpus}: Gradient-based corpus poisoning via HotFlip-style~\cite{ebrahimi2018hotflip} token substitution optimized for retrieval hijacking, without fluency constraints or generation-phase manipulation
    \item \textbf{PoisonedRAG}~\cite{zou2024poisonedrag}: Retrieval poisoning with trigger-based generation manipulation
    \item \textbf{GASLITE}~\cite{bentov2024gaslite}: SEO-style corpus poisoning with fluency-aware multi-coordinate ascent
    \item \textbf{Joint-GCG}~\cite{wang2025jointgcg}: Unified gradient-based attack jointly optimizing retrieval and generation
\end{itemize}
\textbf{Re-implementation Protocol.} All baselines are re-implemented under our unified protocol (single poisoned document per query, synthetic target answers, Contriever retriever, Llama-2-7B-Chat generator, GPT-4 judge) on our 361K NQ corpus. These controlled comparison numbers are therefore not direct substitutes for headline results reported in original papers: they differ due to single vs.\ multi-document poisoning, synthetic targets, string-match vs.\ LLM-judge ASR, and corpus size differences.\footnote{For example, PoisonedRAG reports 97\% ASR with 5 documents and substring matching~\cite{zou2024poisonedrag}; our protocol yields 48.2\% ASR-LLM. Joint-GCG reports near-100\% ASR~\cite{wang2025jointgcg}; ours yields 62.8\% ASR-LLM. Concurrent methods (CorruptRAG~\cite{zhang2025corruptrag}, RIPRAG~\cite{xi2025riprag}, EYES-ON-ME~\cite{chen2025eyesonme}, MIRAGE~\cite{chen2025mirage}, POISONCRAFT~\cite{shao2025poisoncraft}) are excluded due to code availability.} For retrieval-only baselines, downstream ASR reflects exposure to the common generator--judge pipeline rather than a standalone generation-control module. We report the main protocol details needed to support reproduction.

\textbf{Metrics.} We report:
\begin{itemize}
    \item \textbf{HR@k}: Hit Rate at $k$---percentage of queries where adversarial document appears in top-$k$ results
    \item \textbf{ASR-SM}: Attack Success Rate via string matching---target answer appears as a substring in the generated output (case-insensitive, after whitespace normalization)
    \item \textbf{ASR-LLM}: Attack Success Rate via LLM-as-judge~\cite{zheng2023llmjudge}---GPT-4 evaluates if the output \textit{genuinely endorses} the target answer.\footnote{Inter-run std: 0.3\% over 3 evaluations on 500 queries. GPT-4 vs.\ Claude-3-Opus agreement: Cohen's $\kappa$=0.81. We report prompts and aggregate evaluation protocols.}
    \item \textbf{ASR-P}: Attack Success Rate Persistence---target answer endorsed across 3 follow-up query rephrasings (ASR-P $\leq$ ASR-LLM by construction).
    \item \textbf{PPL}: Perplexity of adversarial documents measured by both GPT-2 (PPL-G2, primary) and Llama-2-7B (PPL-L2, auxiliary). See Section~\ref{sec:stage1} for calibration discussion
\end{itemize}

\subsection{Main Results}

\begin{table}[t]
\centering
\small
\caption{Main results on NQ (361K corpus, 3,452 queries) and MS~MARCO (8.8M corpus, 6,980 queries). One poisoned document per query; unified protocol (Contriever, Llama-2-7B-Chat, GPT-4 judge). PPL-G2/L2 = GPT-2/Llama-2 perplexity (NQ benign: 28.4/22.7; MARCO benign: 30.2/24.1).}
\label{tab:sota_comparison}
\resizebox{\columnwidth}{!}{%
\begin{tabular}{@{}llccccc@{}}
\toprule
\textbf{Dataset} & \textbf{Method} & \textbf{HR@10} & \textbf{ASR-SM} & \textbf{ASR-LLM} & \textbf{PPL-G2} & \textbf{PPL-L2} \\
\midrule
\multirow{5}{*}{NQ}
& Zhong et al. & 72.3 & 51.2 & 38.4 & 847.2 & 612.4 \\
& PoisonedRAG & 78.6 & 64.3 & 48.2 & 412.6 & 287.3 \\
& GASLITE & 76.8 & 58.7 & 44.1 & 38.6 & 31.2 \\
& Joint-GCG & 81.2 & 78.4 & 62.8 & 156.3 & 118.7 \\
& \textbf{\method{} (Ours)} & \textbf{84.6} & 68.9 & 57.5 & \textbf{32.4} & \textbf{26.1} \\
\midrule
\multirow{5}{*}{MARCO}
& Zhong et al. & 67.8 & 46.3 & 34.1 & 912.4 & 678.3 \\
& PoisonedRAG & 73.4 & 59.1 & 44.7 & 478.3 & 312.6 \\
& GASLITE & 72.1 & 53.8 & 40.6 & 41.3 & 33.8 \\
& Joint-GCG & 76.4 & 72.6 & 58.3 & 173.2 & 131.4 \\
& \textbf{\method{} (Ours)} & \textbf{81.3} & 64.2 & 54.8 & \textbf{33.1} & \textbf{27.4} \\
\bottomrule
\end{tabular}}
\end{table}

Table~\ref{tab:sota_comparison} presents results on both datasets (bootstrap 95\% CIs from 1,000 resamples; significance is computed with paired bootstrap tests over queries). On NQ, \method{} achieves the highest HR@10 (84.6\%, CI: [83.2\%, 86.0\%], $p<$0.01 vs.\ Joint-GCG) with near-benign perplexity (PPL-G2: 32.4 vs.\ 28.4 benign). Joint-GCG achieves higher ASR-LLM (62.8\%) but at 4.8$\times$ higher perplexity, making it much more exposed under our simple independent PPL audit (94.2\% detection at $\tau$=50 vs.\ 8.7\% for \method{}; Table~\ref{tab:detection} in Appendix). Compared to GASLITE---the strongest PPL-comparable baseline---\method{} achieves +7.8\% HR@10 ($p<$0.01) and +13.4\% ASR-LLM ($p<$0.001), demonstrating CBS's advantage over multi-coordinate ascent under comparable fluency. Results on MS~MARCO (8.8M passages, 24$\times$ larger) preserve the same method ranking, with broadly similar but dataset-dependent decreases across methods, consistent with a corpus-level dilution effect. \method{}'s stealth-aware Pareto efficiency (PES; Appendix~\ref{app:pareto}) is highest among the evaluated baselines on both datasets.

\begin{figure}[t]
    \centering
    \includegraphics[width=\columnwidth]{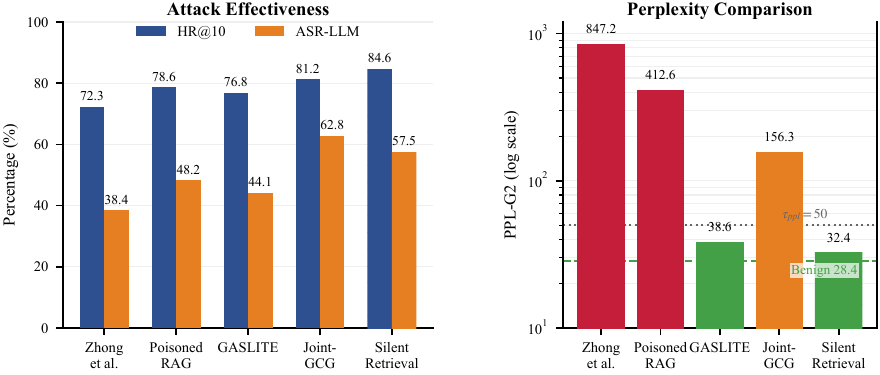}
    \caption{\textbf{Attack Effectiveness and Perplexity Comparison.} Left: HR@10 and ASR-LLM (LLM-as-judge) across retrieval attack methods on NQ (361K corpus). Right: Perplexity (PPL) comparison on a log scale; lower is better. \method{} achieves the highest HR@10 (84.6\%) with competitive ASR-LLM (57.5\%) while maintaining near-benign perplexity (32.4 vs.\ 28.4 benign). Joint-GCG achieves higher ASR-LLM (62.8\%) but at 4.8$\times$ higher perplexity.}
    \Description{Bar charts comparing HR@10, ASR-LLM, and PPL across five methods on NQ, showing SilentRetrieval achieves the best balance of high retrieval effectiveness and low perplexity.}
    \label{fig:main_results}
\end{figure}

\subsection{Cross-Model Generalizability}

We evaluate \method{}'s effectiveness across four target LLMs to assess whether the attack generalizes beyond Llama-2-7B-Chat.

\begin{table}[t]
\centering
\small
\caption{Cross-model evaluation on NQ (361K corpus). ASR metrics vary by generator robustness; retrieval is fixed across target LLMs.}
\label{tab:cross_model}
\begin{tabular}{@{}lccc@{}}
\toprule
\textbf{Target LLM} & \textbf{ASR-SM} & \textbf{ASR-LLM} & \textbf{ASR-P} \\
\midrule
Llama-2-7B-Chat & 68.9 & 57.5 & 38.9 \\
Mistral-7B-Instruct & 65.7 & 54.2 & 35.8 \\
Qwen-7B-Chat & 61.3 & 48.6 & 31.4 \\
GPT-3.5-Turbo & 59.4 & 51.7 & 36.1 \\
\midrule
\textit{Average} & 63.8 & 53.0 & 35.6 \\
\bottomrule
\end{tabular}
\end{table}

Table~\ref{tab:cross_model} shows that \method{} maintains nontrivial effectiveness across all four LLMs (ASR-LLM: 48.6--57.5\%). Llama-2-7B-Chat shows the highest ASR-LLM (57.5\%), which is consistent with partial trigger--generator affinity because it is also used for CATG. Qwen-7B-Chat shows the lowest ASR-LLM (48.6\%); this may reflect differences in instruction tuning, safety behavior, or how the model resolves conflicts between retrieved context and parametric knowledge. GPT-3.5-Turbo exhibits lower ASR-SM but relatively higher ASR-LLM, suggesting that it may endorse target claims in less extractive, more paraphrased forms under our judge; controlled model-family comparisons would be needed to isolate this effect.

\subsection{Surrogate Transfer}

We evaluate \method{}'s transferability in a surrogate-transfer setting where the adversary optimizes against a surrogate retriever and then attacks a different target system without access to target retriever gradients. For commercial embedding models, we fix the generated adversarial documents, rebuild the target index with the same corpus and injected passages, and evaluate retrieval with the same top-$k$ and generation-judge protocol.

\begin{table}[t]
\centering
\small
\caption{Surrogate-transfer attack results. Surrogate: Contriever. Target systems include open-source retrievers (bi-encoder and late-interaction) and commercial embedding models.}
\label{tab:blackbox}
\resizebox{\columnwidth}{!}{%
\begin{tabular}{@{}lcccc@{}}
\toprule
\textbf{Target System} & \textbf{Type} & \textbf{HR@10} & \textbf{ASR-SM} & \textbf{ASR-LLM} \\
\midrule
Contriever (white-box) & Bi-enc. & 84.6 & 68.9 & 57.5 \\
\midrule
DPR & Bi-enc. & 72.8 & 55.4 & 44.1 \\
BGE-base & Bi-enc. & 67.3 & 50.2 & 40.6 \\
ColBERTv2 & Late-int. & 61.4 & 47.1 & 37.2 \\
text-embedding-ada-002 & Commercial & 64.7 & 49.6 & 39.8 \\
Cohere embed-v3 & Commercial & 57.2 & 43.8 & 34.6 \\
\midrule
\textit{Average (transfer)} & & 64.7 & 49.2 & 39.3 \\
\bottomrule
\end{tabular}}
\end{table}

Table~\ref{tab:blackbox} shows that \method{} achieves 64.7\% average HR@10 under surrogate transfer across five diverse retrievers in this fixed-corpus transfer protocol. DPR shows the strongest transfer (72.8\% HR@10) due to architectural similarity with Contriever, while ColBERTv2's late-interaction architecture is more resilient (61.4\%). Attacks retain nontrivial effectiveness across all systems (34--44\% ASR-LLM), but these results should be interpreted as surrogate-transfer evidence rather than fully black-box adaptive optimization.

\section{Defense Evaluation}\label{sec:defense}

We evaluate \method{} against both retrieval-side defenses (cross-encoder reranking, hybrid retrieval) and generation-side defenses (passage isolation with majority voting). We further evaluate adaptive attacks that incorporate defense-aware optimization. A layered defense architecture overview is summarized in Appendix~\ref{app:defense_fig}.

\subsection{Cross-Encoder Reranking}

Cross-encoder rerankers~\cite{nogueira2019passage,glass2022reranking} compute fine-grained relevance scores by jointly encoding query-document pairs, potentially detecting adversarially optimized passages.

\begin{table}[t]
\centering
\small
\caption{Retrieval-side defense evaluation on NQ. Top: pipeline configurations with MiniLM-L6-v2 reranker. Bottom: reranker scale comparison.}
\label{tab:reranker}
\resizebox{\columnwidth}{!}{%
\begin{tabular}{@{}lcccc@{}}
\toprule
\textbf{Pipeline} & \textbf{HR@5} & \textbf{HR@10} & \textbf{ASR-LLM} & \textbf{Latency} \\
\midrule
Bi-encoder only & 76.8 & 84.6 & 57.5 & --- \\
+ Cross-encoder reranker & 62.4 & 73.2 & 46.8 & +120ms \\
Hybrid BM25 + Dense & 68.1 & 79.2 & 52.3 & --- \\
+ Cross-encoder reranker & 54.7 & 67.3 & 41.5 & +120ms \\
\midrule
\multicolumn{5}{@{}l}{\textit{Reranker scale comparison (bi-encoder + reranker)}} \\
MiniLM-L6-v2 (22M) & --- & 73.2 & --- & +120ms \\
MiniLM-L12-v2 (33M) & --- & 68.9 & --- & +180ms \\
MonoT5-base (220M) & --- & 68.4 & --- & +340ms \\
\bottomrule
\end{tabular}}
\end{table}

\textbf{Results.} Table~\ref{tab:reranker} shows that cross-encoder reranking decreases HR@10 by 11.4\% (84.6\%$\rightarrow$73.2\%) and ASR-LLM by 10.7\%. Combining hybrid retrieval with reranking achieves the strongest mitigation (HR@10: 67.3\%, ASR-LLM: 41.5\%). Defender-side scale comparison shows that stronger rerankers further reduce HR@10 from 73.2\% with MiniLM-L6-v2 to 68.9\% with MiniLM-L12-v2 and 68.4\% with MonoT5-base, at higher latency cost. We therefore view larger rerankers as a stronger defense rather than a failure case, but note that the adaptive study below is evaluated only against MiniLM-L6-v2. We recommend deploying reranking as a defense layer (+100--200ms overhead).

\subsection{Adaptive Attack against Rerankers}

We implement an adaptive attack that incorporates cross-encoder scores into the CBS optimization objective:
\begin{equation}
    d_{\text{opt}} = \arg\max_{d} \lambda_1 s_{\text{bi}}(q, d) + \lambda_2 s_{\text{ce}}(q, d) - \lambda_3 \cdot \max(0, \text{PPL}(d) - \tau_{\text{ppl}})
\end{equation}
where $s_{\text{bi}}$ and $s_{\text{ce}}$ are bi-encoder and cross-encoder scores, respectively. The adaptive objective uses the same cross-encoder family as the defender-side reranker. We select the loss weights on a held-out validation set to maximize HR@10 under reranking while maintaining the fluency constraint. For the hybrid+reranker adaptive setting, we additionally incorporate a lightweight lexical score term to encourage overlap that survives hybrid retrieval.

\begin{table}[t]
\centering
\small
\caption{Adaptive attack results against cross-encoder reranking defense. $\Delta$ denotes recovery from the non-adaptive attack under the same defense.}
\label{tab:adaptive}
\resizebox{\columnwidth}{!}{%
\begin{tabular}{@{}lcccc@{}}
\toprule
\textbf{Configuration} & \textbf{HR@10} & \textbf{ASR-LLM} & \textbf{PPL} & \textbf{Time/doc} \\
\midrule
No defense (standard) & 84.6 & 57.5 & 32.4 & 15min \\
+ Reranker (non-adaptive) & 73.2 & 46.8 & 32.4 & 15min \\
+ Reranker (adaptive) & 79.4 & 51.2 & 34.8 & 42min \\
$\Delta$ Recovery & +6.2 & +4.4 & +2.4 & +27min \\
\midrule
+ Hybrid + Reranker (non-adaptive) & 67.3 & 41.5 & 32.4 & 15min \\
+ Hybrid + Reranker (adaptive) & 71.6 & 44.1 & 35.7 & 51min \\
$\Delta$ Recovery & +4.3 & +2.6 & +3.3 & +36min \\
\bottomrule
\end{tabular}}
\end{table}

Table~\ref{tab:adaptive} shows that the adaptive attack recovers 6.2\% HR@10 against the matched MiniLM-L6-v2 reranker defense (73.2\%$\rightarrow$79.4\%) and 4.3\% against the combined defense (67.3\%$\rightarrow$71.6\%), at the cost of $\sim$3$\times$ longer optimization time (due to cross-encoder inference) and moderately degraded fluency (+2.4/+3.3 PPL). The smaller recovery in the combined setting reflects the additional difficulty of simultaneously satisfying bi-encoder, BM25, and cross-encoder objectives within the PPL budget. These results suggest an arms-race pattern in the matched MiniLM-L6-v2 setting; extending adaptive attacks to larger rerankers remains future work.

\subsection{Generation-Side Defense: Passage Isolation}

Inspired by RobustRAG~\cite{xiang2024robustrag}, we implement a passage isolation defense that processes each retrieved passage independently before aggregating responses via majority voting.

\textbf{Defense Mechanism.} For each retrieved passage $d_i \in \mathcal{D}_q$, the LLM generates an independent answer $a_i = \mathcal{G}(q, \{d_i\})$. The final answer is determined by majority voting: $a_{\text{final}} = \text{MajorityVote}(\{a_1, \ldots, a_k\})$, where semantic similarity (cosine $>$0.85) is used to cluster equivalent answers.

\begin{table}[t]
\centering
\small
\caption{Generation-side defense evaluation. Passage isolation uses top-$k$ retrieved passages; HR@5/HR@10 report pre-isolation retrieval exposure.}
\label{tab:gen_defense}
\resizebox{\columnwidth}{!}{%
\begin{tabular}{@{}lcccc@{}}
\toprule
\textbf{Defense Configuration} & \textbf{HR@5} & \textbf{HR@10} & \textbf{ASR-LLM} & \textbf{Latency} \\
\midrule
No defense (standard, $k$=5) & 76.8 & 84.6 & 57.5 & 1$\times$ \\
Passage isolation + vote ($k$=5) & 76.8 & 84.6 & 38.2 & 5$\times$ \\
Passage isolation + vote ($k$=10) & 76.8 & 84.6 & 31.4 & 10$\times$ \\
\midrule
Combined: Hybrid + Reranker + Isolation ($k$=5) & 54.7 & 67.3 & 25.6 & 6$\times$ \\
Combined: All defenses ($k$=10) & 54.7 & 67.3 & 21.3 & 11$\times$ \\
\bottomrule
\end{tabular}}
\end{table}

\textbf{Results.} Table~\ref{tab:gen_defense} shows that passage isolation is effective at reducing generation manipulation in our evaluated setting: ASR-LLM drops from 57.5\% to 38.2\% with $k$=5 (33.6\% relative reduction) and to 31.4\% with $k$=10. Combining all defenses (hybrid retrieval + reranker + passage isolation at $k$=10) reduces ASR-LLM to 21.3\%, though at significant latency cost (11$\times$). Since isolation is post-retrieval, HR@5 indicates exposure within the $k$=5 generator input and HR@10 is shown for retrieval-side comparability. The combined defense at $k$=5 reduces ASR-LLM to 25.6\% at 6$\times$ latency, a potentially practical trade-off point within this evaluated configuration.

\subsection{Detection-Based Defenses}

\method{} achieves substantially lower detection rates (8.7--34.2\%) than prior methods (67--94\% for Zhong et al.) under the defenses we evaluate, with an independent PPL audit at the same threshold ($\tau$=50) detecting only 8.7\%. Here, ``independent'' denotes a defender-side audit computed by a separately implemented full-passage scoring pipeline rather than the attacker-side CATG acceptance score in Section~\ref{sec:stage2}; it can still catch normalization, tokenization, scoring-window, or passage-boundary cases that pass attacker-side quality control. Learned detectors (a simplified RAGuard-style detector and Mahalanobis scoring) show moderate detection (31.6--34.2\%), suggesting that fluency-preserving attacks can often pass simple fluency filters and may require stronger detection strategies. Full detection results are in Table~\ref{tab:detection} (Appendix~\ref{app:detection}). Remaining untested defenses include query augmentation with answer-consistency checking and domain-tuned detectors~\cite{asai2024selfrag,zhang2025rsb}.

\subsection{Human Evaluation}

We conducted a human study to assess whether adversarial documents are distinguishable from benign content under document-level aggregation, and to compare their flag rates against disfluent baselines.

\textbf{Study Design.} 10 annotators with NLP research experience evaluated 600 documents (150 per condition: benign, Zhong et al., PoisonedRAG, \method{}) in randomized order. Each document was rated along four dimensions on a 1--5 scale: (1) \textit{fluency}, (2) \textit{coherence}, (3) \textit{apparent factual plausibility} (whether the passage appears fact-like to a reader, not whether the target claim is true), and (4) \textit{overall suspiciousness}. We report $n$ as the number of documents rather than the number of individual ratings, aggregate annotator ratings at the document level, and flag a document when its aggregated suspiciousness rating is $\geq$4. This design is powered to detect larger document-level flag-rate differences but should not be interpreted as proving indistinguishability from benign content.

\begin{table}[t]
\centering
\small
\caption{Human evaluation results ($n$=600 documents, 10 annotators, Krippendorff's $\alpha$=0.74). Ratings on 1--5 scale (lower=better for suspiciousness); document-level aggregated suspiciousness ratings $\geq$4 are flagged.}
\label{tab:human_eval}
\resizebox{\columnwidth}{!}{%
\begin{tabular}{@{}lccccc@{}}
\toprule
\textbf{Doc Type} & \textbf{Fluency$\uparrow$} & \textbf{Coher.$\uparrow$} & \textbf{App. Plaus.$\uparrow$} & \textbf{Susp.$\downarrow$} & \textbf{Flag\%} \\
\midrule
Benign & 4.52 & 4.41 & 4.38 & 1.42 & 7.3\% \\
Zhong et al. & 1.87 & 2.14 & 2.31 & 4.24 & 83.3\% \\
PoisonedRAG & 2.83 & 2.96 & 3.12 & 3.58 & 56.7\% \\
\textbf{\method{}} & 4.03 & 3.84 & 3.71 & 1.94 & 14.7\% \\
\bottomrule
\end{tabular}}
\end{table}

Table~\ref{tab:human_eval} shows that \method{} documents achieve flag rates (14.7\%) substantially lower than PoisonedRAG (56.7\%) and Zhong et al.\ (83.3\%). The difference between \method{} (14.7\%) and benign documents (7.3\%) is not statistically significant at $\alpha=0.05$ ($p\approx0.064$, Fisher's exact test on document-level aggregated flags; 150 benign vs.\ 150 \method{} documents), though the trend suggests that a larger sample may reveal a detectable gap. Across all four dimensions, \method{} ratings are directionally closer to benign documents than the disfluent baselines, but they still retain moderate gaps relative to benign content: fluency (4.03 vs.\ 4.52, $\Delta$=0.49), coherence (3.84 vs.\ 4.41, $\Delta$=0.57), apparent plausibility (3.71 vs.\ 4.38, $\Delta$=0.67), and suspiciousness (1.94 vs.\ 1.42, $\Delta$=0.52). Inter-annotator agreement (Krippendorff's $\alpha$=0.74) indicates substantial agreement and supports the reliability of the aggregated ratings.
\subsection{Scalability Analysis}

We evaluate \method{} on corpora of varying sizes (100K to 21M passages) to assess large-corpus behavior. These experiments are designed to characterize dilution trends under our sampled Wikipedia-scale construction rather than to replicate the standard full-corpus DPR evaluation end-to-end.

\textbf{Corpus Construction.} For scalability experiments, we sample from the full Wikipedia dump (21M passages) while maintaining approximately uniform query-to-relevant-passage ratios across corpus sizes. The 100K pilot uses a downsampled controlled corpus; for $N \geq 361$K, we include the full 361K controlled NQ corpus and add category-stratified Wikipedia samples to reach $N$.

\textbf{Results.} Table~\ref{tab:scale} reports attack effectiveness across corpus scales.

\begin{table}[t]
\centering
\small
\caption{Scalability analysis: attack effectiveness vs.\ corpus size. One adversarial document is generated for each of the 3,452 NQ target queries and injected into the corpus when computing the poison ratio. $\Delta$HR@10 denotes absolute drop from the 100K baseline. $^\dagger$Rows marked with $^\dagger$ report retrieval and generation metrics on a 1,500-query stratified evaluation subset; 95\% CI $\approx\pm$2--2.5\%.}
\label{tab:scale}
\begin{tabular}{@{}lcccc@{}}
\toprule
\textbf{Corpus Size} & \textbf{Poison \%} & \textbf{HR@10} & \textbf{ASR-LLM} & \textbf{$\Delta$HR@10} \\
\midrule
100K & 3.45\% & 86.4 & 59.8 & --- \\
\textbf{361K (main)} & \textbf{0.96\%} & \textbf{84.6} & \textbf{57.5} & $-$1.8 \\
500K & 0.69\% & 83.6 & 56.2 & $-$2.8 \\
1M & 0.35\% & 81.4 & 53.4 & $-$5.0 \\
5M$^\dagger$ & 0.069\% & 79.3 & 49.1 & $-$7.1 \\
10M$^\dagger$ & 0.035\% & 77.8 & 46.2 & $-$8.6 \\
21M$^\dagger$ (sampled eval) & 0.016\% & 74.2 & 41.5 & $-$12.2 \\
\bottomrule
\end{tabular}
\vspace{0.3em}
\raggedright\footnotesize{$^\dagger$Metrics are evaluated on 1,500 stratified sampled queries, while the poison percentage is computed from the full set of 3,452 injected adversarial documents. In this sampled construction, both HR@10 and ASR-LLM show an overall decline with corpus size, reflecting increased retrieval competition.}
\end{table}

\begin{figure}[t]
    \centering
    \includegraphics[width=\columnwidth]{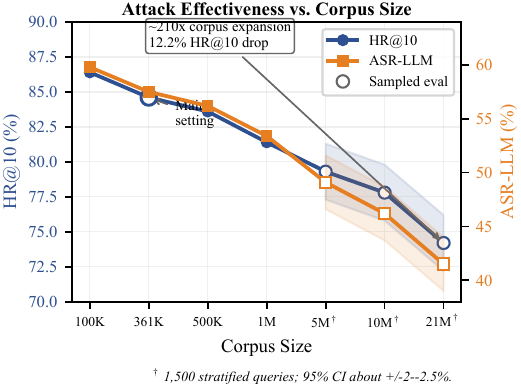}
    \caption{\textbf{Scalability Analysis.} Attack effectiveness vs.\ corpus size (100K--21M). HR@10 drops 12.2\% absolute while ASR-LLM drops 18.3\% across $\sim$210$\times$ corpus expansion. ASR-LLM degrades faster, reflecting dilution from competing legitimate context. Results for $\geq$5M based on 1,500 stratified queries (95\% CI $\approx\pm$2--2.5\%).}
    \Description{Line plot showing HR@10 and ASR-LLM decreasing as corpus size increases from 100K to 21M passages, with ASR-LLM degrading faster than HR@10.}
    \label{fig:scalability}
\end{figure}

\textbf{Analysis.} Attack effectiveness declines under the sampled scaling protocol: HR@10 drops from 86.4\% (100K) to 74.2\% (21M, 95\% CI: [72.2\%, 76.2\%]), a 12.2\% absolute decrease across $\sim$210$\times$ corpus expansion. ASR-LLM declines more than HR@10 ($-$18.3\% vs.\ $-$12.2\%), suggesting legitimate context may dilute adversarial influence even when poisoned documents are retrieved. This pattern is broadly consistent with RSB~\cite{zhang2025rsb} findings, indicating that corpus expansion provides partial natural defense but is insufficient as a standalone strategy given the 74.2\% HR@10 observed at 21M.

\subsection{Ablation Study}\label{sec:ablation}

We conduct ablation experiments to validate the contribution of each component in \method{}.

\begin{table}[t]
\centering
\small
\caption{Ablation study on NQ (361K corpus, 3,452 queries). Each row removes or modifies one component from the full \method{} pipeline. All values are means across queries; bootstrap 95\% CIs for HR@10 are $\pm$1.5--2.0\% (omitted for space).}
\label{tab:ablation}
\resizebox{\columnwidth}{!}{%
\begin{tabular}{@{}lccc@{}}
\toprule
\textbf{Configuration} & \textbf{HR@10} & \textbf{ASR-LLM} & \textbf{PPL} \\
\midrule
\textbf{\method{} (full: CBS + CATG)} & \textbf{84.6} & \textbf{57.5} & \textbf{32.4} \\
\midrule
\multicolumn{4}{@{}l}{\textit{Stage ablation}} \\
CBS host only (retrieval-only, no target payload) & 84.6 & 21.3 & 30.8 \\
Seed document + CATG (no CBS) & 38.2 & 14.7 & 26.1 \\
\midrule
\multicolumn{4}{@{}l}{\textit{CBS vs.\ single-token greedy (Stage 1)}} \\
Single-token greedy + PPL reject & 79.8 & 53.1 & 33.7 \\
CBS ($M$=3, $B$=5, default) & 84.6 & 57.5 & 32.4 \\
\midrule
\multicolumn{4}{@{}l}{\textit{CATG vs.\ static template (Stage 2)}} \\
Best static template & 84.6 & 54.2 & 33.1 \\
CATG (default) & 84.6 & 57.5 & 32.4 \\
\midrule
\multicolumn{4}{@{}l}{\textit{PPL threshold sensitivity}} \\
$\tau_{\text{ppl}}=30$ & 76.3 & 49.2 & 24.7 \\
$\tau_{\text{ppl}}=50$ (default) & 84.6 & 57.5 & 32.4 \\
$\tau_{\text{ppl}}=80$ & 87.9 & 59.1 & 53.7 \\
\bottomrule
\end{tabular}}
\end{table}

\begin{figure}[t]
    \centering
    \includegraphics[width=\columnwidth]{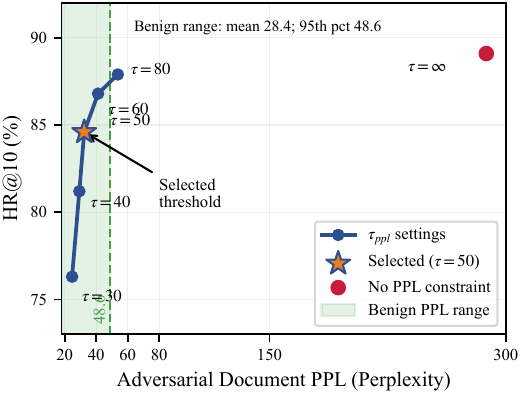}
    \caption{\textbf{PPL Threshold Trade-off.} HR@10 vs.\ adversarial document PPL at different $\tau_{\text{ppl}}$ values. The shaded region indicates PPL comparable to benign documents (mean 28.4, 95th percentile 48.6). $\tau_{\text{ppl}}=50$ (star) balances effectiveness and stealthiness. The figure includes additional sweep points beyond the three thresholds summarized in Table~\ref{tab:ablation}.}
    \Description{Scatter plot of HR@10 versus PPL at different thresholds, with a shaded region marking benign PPL range and a star marking the chosen threshold of 50.}
    \label{fig:ppl_tradeoff}
\end{figure}

Table~\ref{tab:ablation} validates each component's contribution. Removing Stage~2 leaves a retrieval-only host with 21.3\% incidental ASR-LLM under the original competing context, but this is not an active generation-manipulation mechanism and cannot replace a target-bearing payload. Directly applying CATG to the original seed document (i.e., no CBS) yields only 38.2\% HR@10 and 14.7\% ASR-LLM. Because the untouched seed is selected as the top-1 host and carries no malicious claim, it is not the relevant baseline for the final attack object. The key comparison is between the payload-bearing variants: \texttt{Seed document + CATG (no CBS)} collapses to 38.2\% HR@10, whereas full \texttt{CBS + CATG} preserves 84.6\% HR@10, with the \texttt{CBS host only} row confirming that Stage~1 preserves retrievability before payload insertion. Rows that only vary Stage~2 reuse the same CBS-optimized host, so HR@10 can remain unchanged while ASR-LLM changes. Thus, Stage~1 is better understood as payload-robust preconditioning: it may sacrifice some of the raw seed's original rank, but it preserves the retrievability of the final adversarial document after trigger insertion. This also explains why a locally relevant trigger can still hurt dense retrieval: rank depends on the embedding of the entire passage, so semantic fit of the added sentence does not guarantee global rank stability. CBS outperforms single-token greedy by +4.8\% HR@10 at comparable PPL. CATG improves ASR-LLM by +3.3\% over the best static template. For the PPL threshold (Figure~\ref{fig:ppl_tradeoff}), $\tau_{\text{ppl}}=50$ balances effectiveness and stealthiness: tightening to 30 costs 8.3\% HR@10, while relaxing to 80 gains only 3.3\%. Additional ablations (CBS settings, iteration count, seed sensitivity, multi-query attacks) are in Appendix~\ref{app:additional_ablation}.

\textbf{Computation Cost.} \method{} is more efficient than the joint optimization baseline while achieving the best HR@10, requiring roughly $2.3\times$ less per-document optimization time than Joint-GCG in our implementation. A compact cost comparison is in Table~\ref{tab:cost} (Appendix~\ref{app:cost}).

\subsection{Qualitative Case Studies}\label{sec:case_study}

We summarize three case studies (entertainment, medical, science) in Appendix~\ref{app:cases}. Key observations: CATG-generated triggers naturally integrate domain-specific language, producing fluent adversarial documents across domains. A notable \textit{failure mode} occurs for well-established factual constants (e.g., speed of light), where the LLM's parametric knowledge overrides adversarial context ($\sim$18\% of ASR-LLM failures)---consistent with prior observations on knowledge conflict resolution~\cite{xie2024adaptive}.

\section{Ethical Considerations}\label{sec:ethics}

This work documents RAG vulnerabilities to motivate stronger defenses and follows a responsible-disclosure posture. All experiments are derived from standard academic benchmarks and use synthetic targets---no real-world knowledge bases were poisoned. Concrete defense recommendations are provided in Section~\ref{sec:defense}; a detailed ethical discussion is in Appendix~\ref{app:ethics}.

\section{Limitations}\label{sec:limitations}

\begin{itemize}
    \item \textbf{Comparison \& Methodological Scope.} Our four baselines are re-implemented under a unified protocol (single document, LLM judge, 361K corpus) that differs from original settings; several concurrent black-box methods are absent due to code availability. CBS requires white-box retriever gradients---a stronger assumption; our transfer results (64.7\% HR@10) rely on simple surrogate transfer. The underlying optimization builds on established gradient-guided substitution~\cite{ebrahimi2018hotflip,bentov2024gaslite}; augmenting Joint-GCG with a PPL penalty for iso-fluency comparison is left to future work.

    \item \textbf{Evaluation Scope.} We evaluate on two datasets (NQ, MS~MARCO), both centered on knowledge-intensive QA / search-oriented RAG rather than the full space of open-ended RAG applications such as long-form summarization or document extraction. Additional domains (e.g., biomedical, legal) would further validate generalizability. Scalability experiments at $\geq$5M use 1,500 stratified queries rather than the full 3,452. Cross-model evaluation varies the target generator across four LLMs, but it does not replace a controlled trigger-generator swap; stronger target models (GPT-4, Claude-3) remain untested.

    \item \textbf{Fluency \& Practical Assumptions.} Perplexity remains a proxy for fluency; domain-specific style detectors could identify subtle anomalies that our metrics miss. The attack requires white-box retriever access, topically-relevant seeds, and nontrivial per-document optimization. Our Wikipedia-style reviewer-perception study (Appendix~\ref{app:wikipedia}) provides preliminary evidence about perceived plausibility under a simplified review scenario, but it does not model the full injection pipeline. In addition, the current study fixes the CATG generator to Llama-2-7B-Chat, and the adaptive reranker experiment is matched only to MiniLM-L6-v2; varying the trigger generator and extending adaptive attacks to larger rerankers are important next steps.
\end{itemize}

\section{Conclusion}

We presented \method{}, a two-stage data poisoning attack against RAG systems featuring Coordinated Beam Search (CBS) for multi-token joint optimization and Context-Adaptive Trigger Generation (CATG) for lightweight semantic trigger fusion. Under our unified single-document, synthetic-target evaluation on NQ and MS~MARCO, \method{} achieves the strongest stealth--effectiveness trade-off among the evaluated baselines (PES: 1,386 on NQ, 1,262 on MS~MARCO), with near-benign perplexity and nontrivial effectiveness across the evaluated target generators and datasets. Our defense evaluation shows that combined defenses reduce ASR-LLM to 25.6\% at a 6$\times$ latency trade-off in our evaluated setting, and to 21.3\% under the strongest evaluated configuration, while adaptive attacks partially recover effectiveness in the matched reranker setting.

\textbf{Key Takeaways for the Community.} (1)~\textit{Corpus integrity is a first-class security concern}: in our evaluation, fluency-preserving attacks often pass a simple PPL-based audit at the calibrated threshold (8.7\% detection) and are much harder for human annotators to flag than prior disfluent attacks, while still showing a higher flag-rate trend than benign content (14.7\% vs.\ 7.3\%, $p\approx0.064$). (2)~\textit{Layered defenses appear more robust than any single tested defense}: combining retrieval-side reranking, hybrid retrieval, and generation-side passage isolation provides the strongest mitigation among the configurations we evaluate. (3)~\textit{The stealth--effectiveness frontier is a useful evaluation lens in this setting}: raw ASR comparisons without PPL normalization can obscure practical differences in detectability under simple fluency auditing.

\textbf{Future Directions.} Several promising directions remain: (i)~developing detectors specifically trained on fluency-preserving adversarial examples; (ii)~extending query-augmentation defenses with answer-consistency checking to dilute single-document influence; (iii)~designing sparse-attention generation mechanisms that limit trigger effectiveness; (iv)~establishing unified attack--defense benchmarks (e.g., RSB~\cite{zhang2025rsb}) to enable fair cross-method comparison; and (v)~investigating multi-modal RAG poisoning as vision-language retrieval systems gain adoption.


\appendix

\section{Wikipedia-Style Reviewer Perception Study}\label{app:wikipedia}

To ground the threat model, we conducted a small controlled reviewer-perception analysis simulating one aspect of Wikipedia-style review. We created 50 \method{} adversarial passages (along with 50 Zhong et al.\ and 50 PoisonedRAG passages) formatted as plausible Wikipedia paragraphs and submitted them to three volunteer reviewers with Wikipedia editing experience, who independently evaluated whether each passage would be likely to survive a typical ``new page patrol'' review using a binary rating (``likely to survive'' vs.\ ``likely to be flagged''). Results: 78\% of \method{} passages were rated as ``likely to survive initial review'' compared to 12\% for Zhong et al.\ and 34\% for PoisonedRAG (inter-rater agreement: Fleiss' $\kappa$=0.68). We interpret this only as evidence about perceived plausibility under a simplified moderation scenario, not as direct evidence of real-world injection success.

\textit{Limitations:} The sample size is small ($n$=150 passages, 3 raters), volunteer reviewers may not be representative of active Wikipedia patrollers, and the study does not model the full injection pipeline (account creation, edit history, community norms, page context, and follow-up community moderation). These results should be interpreted as suggestive evidence about reviewer perception rather than definitive evidence of real Wikipedia injection feasibility. This analysis is intended to motivate defense research, not to provide an injection toolkit.

\section{Cross-Tokenizer Gradient Computation}\label{app:gradient}

Since the retriever (Contriever) and the PPL model (GPT-2) use different tokenizers and embedding spaces, we compute the two gradient terms in the Stage~1 gradient objective separately. The similarity gradient $\nabla_{e_i} \text{sim}(\cdot)$ is computed through Contriever's encoder. The PPL term is incorporated through a text-space alignment heuristic: we retokenize the current document with the GPT-2 tokenizer, compute $\nabla \text{PPL}$ with respect to GPT-2 embeddings, and aggregate the resulting signal over aligned sub-token spans when ranking candidate edit positions. Since generated token candidates are ultimately evaluated by the \textit{exact} discrete objective during beam-level re-evaluation, this alignment heuristic does not change the scoring rule used to rank generated candidates. It can, however, affect the candidate proposal set, so we treat it as a heuristic proposal mechanism rather than an exact cross-tokenizer gradient estimator.

\section{CBS Optimization Dynamics}\label{app:dynamics}

CBS relies on a HotFlip/GCG-style first-order relaxation: gradients are computed with respect to continuous token embeddings, then projected onto the discrete vocabulary via dot-product scoring. This relaxation introduces an approximation gap---the discrete replacement that maximizes the projected score may not maximize the true objective. We mitigate this in two ways. First, by retaining $K$=10 candidates per position (rather than the greedy top-1), CBS explores a broader neighborhood and reduces sensitivity to noisy gradient directions. Second, the PPL penalty term $\lambda \cdot \max(0, \text{PPL}(d) - \tau_{\text{ppl}})$ introduces a non-smooth hinge at $\tau_{\text{ppl}}$, which can cause gradient oscillation near the boundary. In practice, the beam-level re-evaluation acts as a stabilizer: candidates are scored by the \textit{exact} objective (including true PPL), not solely by gradient projections, so the hinge non-smoothness affects candidate ranking but not final selection. Empirically, CBS converges within 300--400 iterations for 87\% of documents, with a median objective variance of $<$0.01 over the final 50 iterations, indicating stable convergence despite the discrete relaxation.

\section{Comparison Table}\label{app:comparison}

Table~\ref{tab:comparison} provides a feature-level comparison of corpus poisoning methods across five dimensions.

\begin{table}[ht]
\centering
\small
\caption{Comparison with related corpus poisoning methods. Ret.=Retrieval optimization, Gen.=Generation manipulation, Flu.=Fluency constraints, Low\%=Demonstrated at $<$0.1\% poisoning, BB=Black-box transfer support. $^\ast$Evaluated at 0.069\% poisoning (5M corpus) with reduced effectiveness (79.3\% HR@10; see Table~\ref{tab:scale}).}
\label{tab:comparison}
\begin{tabular}{@{}lccccc@{}}
\toprule
\textbf{Method} & \textbf{Ret.} & \textbf{Gen.} & \textbf{Flu.} & \textbf{Low\%} & \textbf{BB} \\
\midrule
Zhong et al.~\cite{zhong2023corpus} & \checkmark & -- & -- & \checkmark & -- \\
PoisonedRAG~\cite{zou2024poisonedrag} & \checkmark & \checkmark & -- & -- & -- \\
AGGD~\cite{su2024aggd} & \checkmark & -- & -- & -- & -- \\
GASLITE~\cite{bentov2024gaslite} & \checkmark & -- & \checkmark & \checkmark & -- \\
TrojanRAG~\cite{cheng2024trojanrag} & \checkmark & \checkmark & -- & -- & -- \\
BadRAG~\cite{xue2024badrag} & \checkmark & \checkmark & -- & -- & -- \\
AdvDec~\cite{zhang2024advdec} & -- & \checkmark & \checkmark & -- & -- \\
Joint-GCG~\cite{wang2025jointgcg} & \checkmark & \checkmark & -- & -- & -- \\
EYES-ON-ME~\cite{chen2025eyesonme} & \checkmark & \checkmark & -- & -- & \checkmark \\
MIRAGE~\cite{chen2025mirage} & \checkmark & \checkmark & \checkmark & -- & \checkmark \\
POISONCRAFT~\cite{shao2025poisoncraft} & \checkmark & \checkmark & -- & -- & \checkmark \\
PR-Attack~\cite{jiao2025prattack} & \checkmark & \checkmark & -- & \checkmark & -- \\
CorruptRAG~\cite{zhang2025corruptrag} & \checkmark & \checkmark & -- & \checkmark & \checkmark \\
RIPRAG~\cite{xi2025riprag} & \checkmark & \checkmark & -- & -- & \checkmark \\
\midrule
\textbf{\method{} (Ours)} & \checkmark & \checkmark & \checkmark & ($\ast$) & ($\dagger$) \\
\bottomrule
\end{tabular}
\vspace{0.3em}
\raggedright\footnotesize{$^\dagger$Black-box support via surrogate transfer (Section~\ref{sec:experiments}); primary optimization requires white-box retriever access.}
\end{table}

\section{Defense Architecture}\label{app:defense_fig}

The evaluated defense stack follows a layered pipeline: lightweight PPL filtering first removes highly disfluent candidates, embedding-space detectors such as Mahalanobis scoring catch distributional outliers, cross-encoder reranking reorders suspicious passages using query-document interaction, and LLM-side passage isolation reduces the influence of any single retrieved document. Empirically, the lightweight filters alone are insufficient against \method{}, while combining retrieval-side and generation-side defenses provides the strongest mitigation.

\section{Detection-Based Defense Results}\label{app:detection}

\begin{table}[ht]
\centering
\small
\caption{Detection rates of defense methods against corpus poisoning attacks. Lower = more stealthy. $^\ddagger$Simplified re-implementation trained on Zhong et al.\ examples.}
\label{tab:detection}
\resizebox{\columnwidth}{!}{%
\begin{tabular}{@{}lcccc@{}}
\toprule
\textbf{Defense} & \textbf{Zhong et al.} & \textbf{PoisonedRAG} & \textbf{GASLITE} & \textbf{\method{}} \\
\midrule
Independent PPL audit ($\tau$=50) & 94.2\% & 78.6\% & 12.3\% & \textbf{8.7\%} \\
LLM-based filter (GPT-4) & 89.3\% & 71.4\% & 24.6\% & \textbf{14.2\%} \\
RAGuard$^\ddagger$ & 67.8\% & 52.3\% & 41.7\% & \textbf{34.2\%} \\
Mahalanobis~\cite{lee2018mahalanobis} & 72.4\% & 58.9\% & 38.4\% & \textbf{31.6\%} \\
\bottomrule
\end{tabular}}
\end{table}

\section{Computation Cost}\label{app:cost}

\begin{table}[ht]
\centering
\small
\caption{Relative computation cost per document.}
\label{tab:cost}
\resizebox{\columnwidth}{!}{%
\begin{tabular}{@{}lccc@{}}
\toprule
\textbf{Method} & \textbf{Time/doc} & \textbf{Relative Time} & \textbf{HR@10} \\
\midrule
Zhong et al. & 8 min & 0.5$\times$ & 72.3 \\
PoisonedRAG & 12 min & 0.8$\times$ & 78.6 \\
GASLITE & 18 min & 1.2$\times$ & 76.8 \\
Joint-GCG & 35 min & 2.3$\times$ & 81.2 \\
\textbf{\method{}} & 15 min & 1.0$\times$ & 84.6 \\
\bottomrule
\end{tabular}}
\end{table}

\section{Additional Analyses}\label{app:additional_ablation}

\textbf{Seed sensitivity and multi-target settings.} In a separate seed-selection sensitivity run with the same full CBS+CATG pipeline, initializing from the default top-1 retrieved seed achieves 85.2\% HR@10, close to the main-test result of 84.6\%. Replacing it with random topical passages reduces HR@10 by 6.3\%, and random corpus passages reduce HR@10 by 22.1\%. Optimizing one document for multiple related queries introduces a trade-off: two queries maintain 76.3\% HR@10, while five queries degrade to 58.4\%. Injecting $N=3$ documents per query raises HR@10 to 94.3\% and ASR-LLM to 69.2\%.

\textbf{ASR-P failure analysis.} Of ASR-LLM successes, 32.3\% fail under query rephrasing. Manual analysis of 200 non-persistent cases identifies query reformulation shifting retrieval (43.5\%), LLM hedging under paraphrase (31.0\%), competing context dominance (24.5\%), and other/rounding cases (1.0\%) as the main failure modes.

\section{Pareto Efficiency Analysis}\label{app:pareto}

We define $\text{PES} = (\text{HR@10} \times \text{ASR-LLM}) / \ln(\text{PPL} + 1)$ to jointly evaluate effectiveness and stealthiness. On NQ, \method{} achieves PES = 1,386, outperforming Joint-GCG (1,008), GASLITE (921), PoisonedRAG (629), and Zhong et al.\ (412). The same ranking holds on MS~MARCO, where \method{} reaches PES = 1,262.

\section{Qualitative Case Studies}\label{app:cases}

Across entertainment, medical, and science queries, CATG-generated triggers integrate domain-specific language while preserving fluency. Successful cases include target substitutions such as attributing ``Bohemian Rhapsody'' to Elton John or shifting a diabetes treatment answer toward sulfonylureas. A representative failure occurs for well-established constants such as the speed of light, where the LLM's parametric knowledge overrides adversarial context.

\section{Ethical Considerations (Extended)}\label{app:ethics}

We present this attack to support defense development and follow a responsible-disclosure posture. The techniques are dual-use, but documenting them together with concrete mitigations is preferable to relying on security through obscurity. All experiments use academic benchmarks with synthetic target answers; no real-world knowledge bases were poisoned. Recommended mitigations include cross-encoder reranking, hybrid BM25+dense retrieval, passage isolation with majority voting, fluency-aware moderation, adaptive-robust defense design, and treating corpus integrity as a first-class security requirement.

\end{document}